\documentclass[prb,twocolumn,superscriptaddress]{revtex4-1}
\usepackage{graphicx}% Include figure files
\usepackage{dcolumn}% Align table columns on decimal point
\usepackage{bm}% bold math
\usepackage{textgreek}
\usepackage{textcomp}
\usepackage[usenames]{color}

\begin{document}

\title{Localized segregation of gold in ultra-thin Fe films on Au(001)}
\author{P. Gospodari\v{c}} 
%\affiliation{Peter Gr\"{u}nberg Institut, PGI-6, Forschungszentrum J\"ulich GmbH, D-52425 J\"ulich, Germany}
\affiliation{PGI-6, Forschungszentrum J\"ulich GmbH, D-52425 J\"ulich, Germany}
\author{E. M\l y\'{n}czak}
\email{e.mlynczak@fz-juelich.de}
\affiliation{PGI-6, Forschungszentrum J\"ulich GmbH, D-52425 J\"ulich, Germany}
\affiliation{Faculty of Physics and Applied Computer Science, AGH University of Science and Technology, al. Mickiewicza 30, 30-059 Krakow, Poland}
\author{M. Eschbach}
\affiliation{PGI-6, Forschungszentrum J\"ulich GmbH, D-52425 J\"ulich, Germany}
\author{M. Gehlmann}
\affiliation{PGI-6, Forschungszentrum J\"ulich GmbH, D-52425 J\"ulich, Germany}
\author{G. Zamborlini}
\affiliation{PGI-6, Forschungszentrum J\"ulich GmbH, D-52425 J\"ulich, Germany}
\author{V. Feyer}
\affiliation{PGI-6, Forschungszentrum J\"ulich GmbH, D-52425 J\"ulich, Germany}
\author{L. Plucinski}
\affiliation{PGI-6, Forschungszentrum J\"ulich GmbH, D-52425 J\"ulich, Germany}
\author{C. M. Schneider}
\affiliation{PGI-6, Forschungszentrum J\"ulich GmbH, D-52425 J\"ulich, Germany}

\date{\today}

\begin{abstract}
The growth of up to 10 monolayer-thick Fe films on a Au(001) surface was investigated during deposition at room temperature and during annealing, using low-energy electron diffraction and x-ray photoemission spectroscopy as well as locally with low-energy electron microscopy and photoemission electron microscopy. The growth proceeds with a submonolayer of Au segregating to the surface of Fe, which is in agreement with previous studies. Annealing was found to be critical for the presence of Au on the Fe surface. Our study identifies a spatially-inhomogeneous Au segregation mechanism, which proceeds by the formation of cracks in the Fe film, starting at the annealing temperature of 190~$^{\circ}$C, through which Au diffuses towards the surface. As a result, a system with a non-uniform surface electronic structure is obtained. This study shows the necessity to employ spatially-resolved techniques to fully understand the growth modes of the layered epitaxial systems.
\end{abstract}

\maketitle

%%%%%%%%%%%%%%%%%%%%%%%%%%%%%%%%%%%%%%%%%%%%%%%%%%%%%%%%%%%%%%%%%%%%%%%%%%%%%%%%%%%%%%%%%%%%%%
%%%%%%%%%%%%%%%%%%%%%%%%%%%%%%%%%%%%%%%%%%%%%%%%%%%%%%%%%%%%%%%%%%%%%%%%%%%%%%%%%%%%%%%%%%%%%%
\section{Introduction}
The coexistence of exchange and spin-orbit interaction at the ferromagnet/heavy metal interface leads to a torque on the magnetization of the ferromagnet, when a current is applied parallel to the layers.\cite{Miron2010, Garello2013} This recent finding stimulated the revival of fundamental research on the electronic properties of the ferromagnets interfaced with heavy metals.\cite{Moras2015, Carbone2016, Mlynczak2016} One of the possible candidates for such a model system is an Fe film grown on Au(001). Ultra-thin Fe films grown epitaxially on Au(001) have been widely studied because of the very small lattice mismatch $<$ 1 \% and epitaxial growth conditions.\cite{Bonell2013} Thus, it was often the system of choice for studies of  ultra-thin film phenomena, such as quantum well states \cite{Himpsel1991, Ortega1993} and interlayer exchange coupling.\cite{Ortega1993, Shintaku1993, Unguris1994} In a recent publication,\cite{Mlynczak2016} we demonstrated that in the Fe/Au(001) system it is possible to observe the opening/closing of the magnetization-dependent spin-orbit gaps located near the Fermi level. Clearly, Fe/Au(001) system is well-suited as a model ferromagnet/heavy metal system. A challenge that it poses, however, is the sharpness of the  Fe-Au interface as both metals have a tendency towards intermixing.  
\par
The results of several studies of Fe/Au(001) growth were summarized by Bonell \textit{et al.}.\cite{Bonell2013} Various surface-sensitive techniques were applied to determine the growth mode. The majority of studies agree that one monolayer (ML) of Au remains on the surface of 15-20~ML Fe films when deposited at room temperature (RT). Because the Au overlayer lowers the surface free energy of Fe, it is considered to promote the layer-by-layer growth of the first Fe MLs. However, due to different experimental conditions there are a lot of discrepancies about the existence of a Au ML on the surface for thicker Fe films.\cite{Bonell2013} The Au overlayer leads to important changes in the properties of the Fe films. For example, calculations show that it is the Au monolayer which reduces the Fe magnetic moment as compared to the value for Fe/MgO(001).\cite{Li1988, Wu1994} Moreover, the overlayer of Au was suggested to affect the magnetic anisotropy of the Fe film,\cite{Liu1989} which was further supported by theoretical studies.\cite{Guo1991, Szunyogh1995, Gallego2004}
\par
In this paper we study the growth of thin Fe films (up to 10~ML) on a Au(001) single crystal using a combination of spatially resolved experimental techniques: synchrotron-based photoemission electron microscopy (PEEM) and low-energy electron microscopy (LEEM). These two methods offer chemical and structural sensitivity, respectively, with below micrometer-scale resolution. The measurements were carried out at the Nanospectroscopy beamline of the Elettra storage ring (Trieste, Italy). The beamline endstation is equipped with a SPELEEM III microscope (Elmitec, GmbH),\cite{Mentes2014, Locatelli2011} which combines LEEM and energy-filtered x-ray PEEM (XPEEM). Furthermore, the design of the experimental set-up allows imaging at the detector both the back-focal and the analyzer dispersion plane. By inserting a micrometer-size aperture it is possible to limit the probed area to a few \textmu m\textsuperscript{2} and acquire microprobe low energy diffraction (\textmu-LEED) patterns and microprobe x-ray photoemission spectra (\textmu-XPS). The geometry of the microscope \cite{Schmidt1998} gives the opportunity to image or spectroscopically study in-situ deposition of materials under ultra-high vacuum (UHV) conditions during the growth and annealing procedure. 

\par
We start with the analysis of the clean and reconstructed Au(001) surface, for which we analyze the size of the reconstruction domains using LEEM (Sec. \ref{Sec1}). Further, we discuss Fe growth which we monitored by collecting LEED patterns and XPS spectra during the deposition (Sec. \ref{Sec2}). In Sec. \ref{Sec3} we present a combination of LEEM and XPEEM images acquired during annealing, which reveal that Au segregates towards the Fe surface in a non-uniform way. We suggest a possible mechanism for the segregation of Au to the surface of a 10~ML Fe film grown in oblique deposition geometry. 
\begin{figure*}
	
	\includegraphics[width=0.9\textwidth]{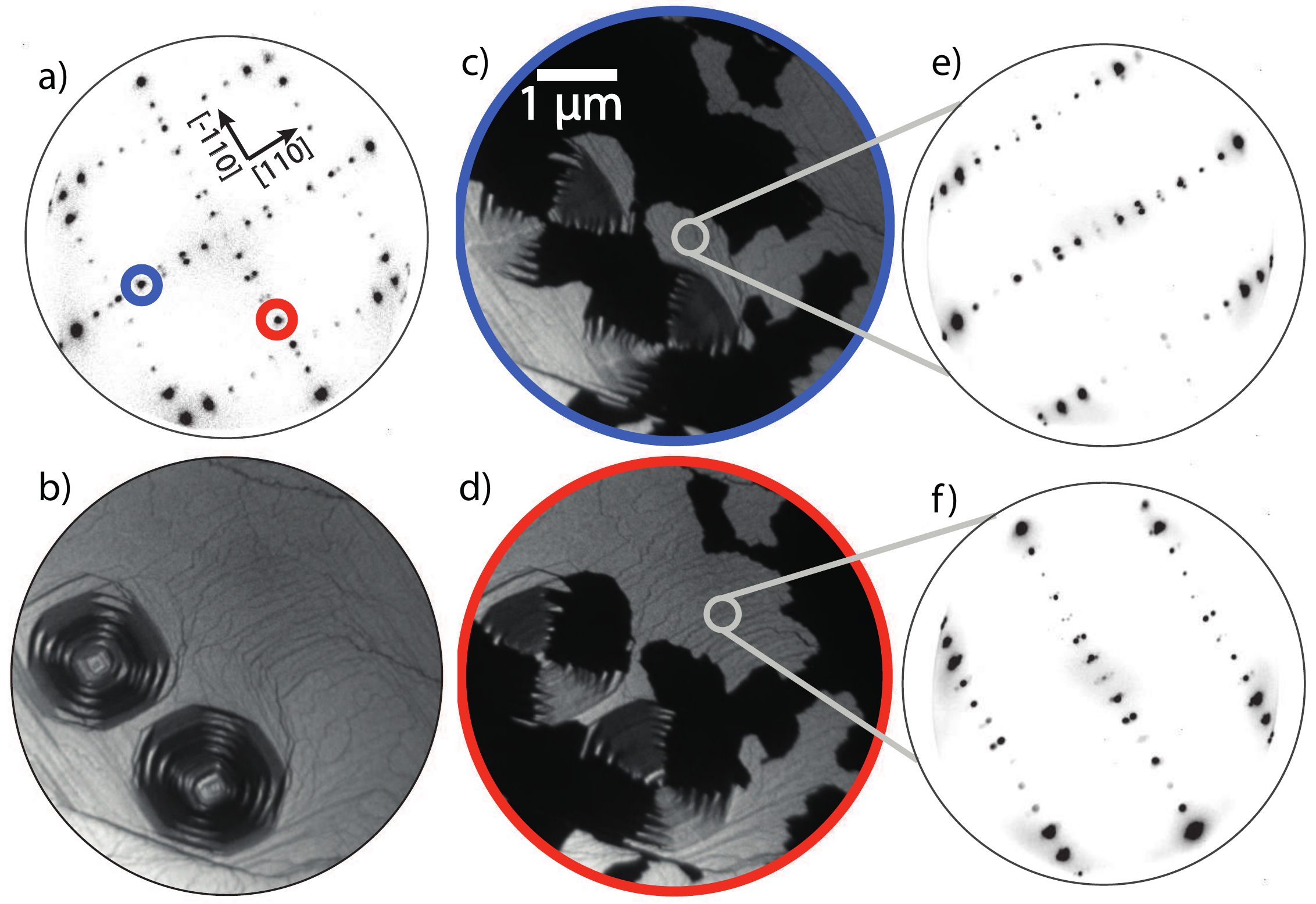}
	\caption{\label{fig:fig1} a) The LEED pattern and b) a LEEM image of the clean Au(001) single crystal surface with two surface defects. In c) and d) the dark-field LEEM images are shown, which were acquired with the contrast aperture closed to the LEED spots marked in (a). The orthogonal domains of the Au(001) surface reconstruction can be clearly distinguished. The \textmu-LEED patterns obtained from the single domain areas are shown in e) and f). The LEED patterns were obtained at 50~eV and the LEEM images at 12~eV electron kinetic energy.}
\end{figure*}
\section{Preparation and characterization of A\MakeLowercase{u}(001)}
\label{Sec1}

Prior to the Fe film deposition, the surface of the Au(001) single crystal was prepared under UHV conditions by multiple cycles of sputtering (Ar$ + $ ions, 30~min, 1.5~keV) and subsequent annealing at 470~$^{\circ}$C. 
\par
The surface quality was characterized by means of LEED and LEEM. The LEED pattern (Fig. 1 (a)) shows the typical reconstruction of the Au(001) crystal face,\cite{Spiridis1999} which was recently refined as c(28x48).\cite{Hammer2014} The presence of the surface reconstruction indicates a clean and well-ordered surface. In the LEEM image (Fig. 1 (b)) the surface steps and the pyramidal defects of the Au(001) surface can be distinguished. It was shown that such structures can form during the indentation of the Au(001) surface to relieve the elastic energy.\cite{Gannepalli2002} Here they are possibly a consequence of the sputtering procedure. 

Two orthogonal rotational domains of the surface reconstruction were observed in the LEEM images (Fig. 1 (c) and (d)). By closing the contrast aperture \cite{Schmidt1998} to select a fractional order diffraction spot, it is possible to laterally resolve a given surface phase. This imaging mode is known as dark-field (DF) LEEM. Two fractional spots were selected, one along the Au[1$\bar{1}$0] and one along the Au[$\bar{1}\bar{1}$0] axis (marked with red and blue circles in Fig. 1 (a)) and the resulting images are shown in Fig. 1 (c) and (d). By comparing these two images it is evident that the two Au surface phases are complementary (one image is the negative of the other). By acquiring \textmu-LEED patterns from two areas with different reconstructions (Fig 1 (e) and (f)) we can separate the contributions to the laterally averaged diffraction pattern (Fig 1 (a)). It is immediately visible that the two domains are orthogonal with respect to each other. We observed no patches of an unreconstructed surface.
The pyramidal surface defects have a clear contrast in the DF LEEM images (Fig. 1 (d) and (e)) as observed earlier by Bauer.\cite{Bauer1994} The corresponding LEED patterns suggest that the terraces of the defects are reconstructed and that the orientation of the steps determines the direction of the surface reconstruction.

%%%%%%%%%%%%%%%%%%%%%%%%%%%%%%%%%%%%%%%%%%%%%%%%%%%%%%%%%%%%%%%%%%%%%%%%%%%%%%%%%%%%%%%%%%%%%%
%%%%%%%%%%%%%%%%%%%%%%%%%%%%%%%%%%%%%%%%%%%%%%%%%%%%%%%%%%%%%%%%%%%%%%%%%%%%%%%%%%%%%%%%%%%%%%
\section{F\MakeLowercase{e} THIN FILM GROWTH ON A\MakeLowercase{u}(001) SURFACE}
\label{Sec2}

The Fe films were grown at RT on the Au(001) substrate using molecular beam epitaxy. For the Fe evaporator a calibration on W(110) single crystal was used. The deposition direction was at a 74$^{\circ}$ angle to the sample surface normal along the Fe[$\bar{1}\bar{1}$0] direction (marked in Fig. 2 (c)). 
\par
During growth the LEED patterns were recorded using electrons with kinetic energy of 50~eV. The patterns obtained at different deposition times and the corresponding angular profiles of the 0-order diffraction spot are shown in Fig. 2 (a). 
After the deposition of 0.3 - 0.4~ML  of Fe the Au(001) surface reconstruction is lifted, which is in agreement with previous studies.\cite{Opitz1997, Blum1999} For an Fe coverage above 1~ML the characteristic LEED pattern of the body-centered-cubic (bcc) Fe lattice appears according to the epitaxial relation: Fe[100]$ \parallel $Au[110] and Fe[010]$ \parallel $Au[1$\bar{1}$0]. The Fe LEED pattern remains well defined at all studied Fe film thicknesses. On the other hand, the LEED~(0,0) spot angular profile changes with increasing Fe film thickness, as demonstrated in Fig. 2 (b). Below 1~ML coverage a sharp peak is observed, while for the coverage above 1~ML the peak shape can be decomposed into a sharp central peak accompanied by a broad background intensity. The different angular profile of the LEED~(0,0) spot at increasing thickness may be related to a change in the interlayer distance during growth, which was observed in other studies,\cite{Opitz1997, Begley1993, Blum1999, Hernan1998} and probably to increasing surface roughness with higher coverage. At half~ML coverage no additional modulation of the already mentioned gradual increase of the diffuse background in the (0,0) spot profile was observed. At Fe coverage of 3~ML the angular profile starts changing into a broad peak which remains broad at 4-10~ML. This may be due to increased roughness of the surface and/or chemical inhomogeneity of the surface (Au-Fe intermixing), which modifies the scattering conditions. Furthermore, LEEM images were acquired during the Fe film growth using electron kinetic energy of 12~eV. Up to 8~ML Fe coverage the distinct step bunches on the surface of the Au single crystal remain visible with LEEM (Fig 2 (c)). For 10~ML we observe a homogeneous intensity of the LEEM image, where the only distinguishable features are the pyramidal defects.     
\par
\begin{figure} [h]

\includegraphics[width=\columnwidth]{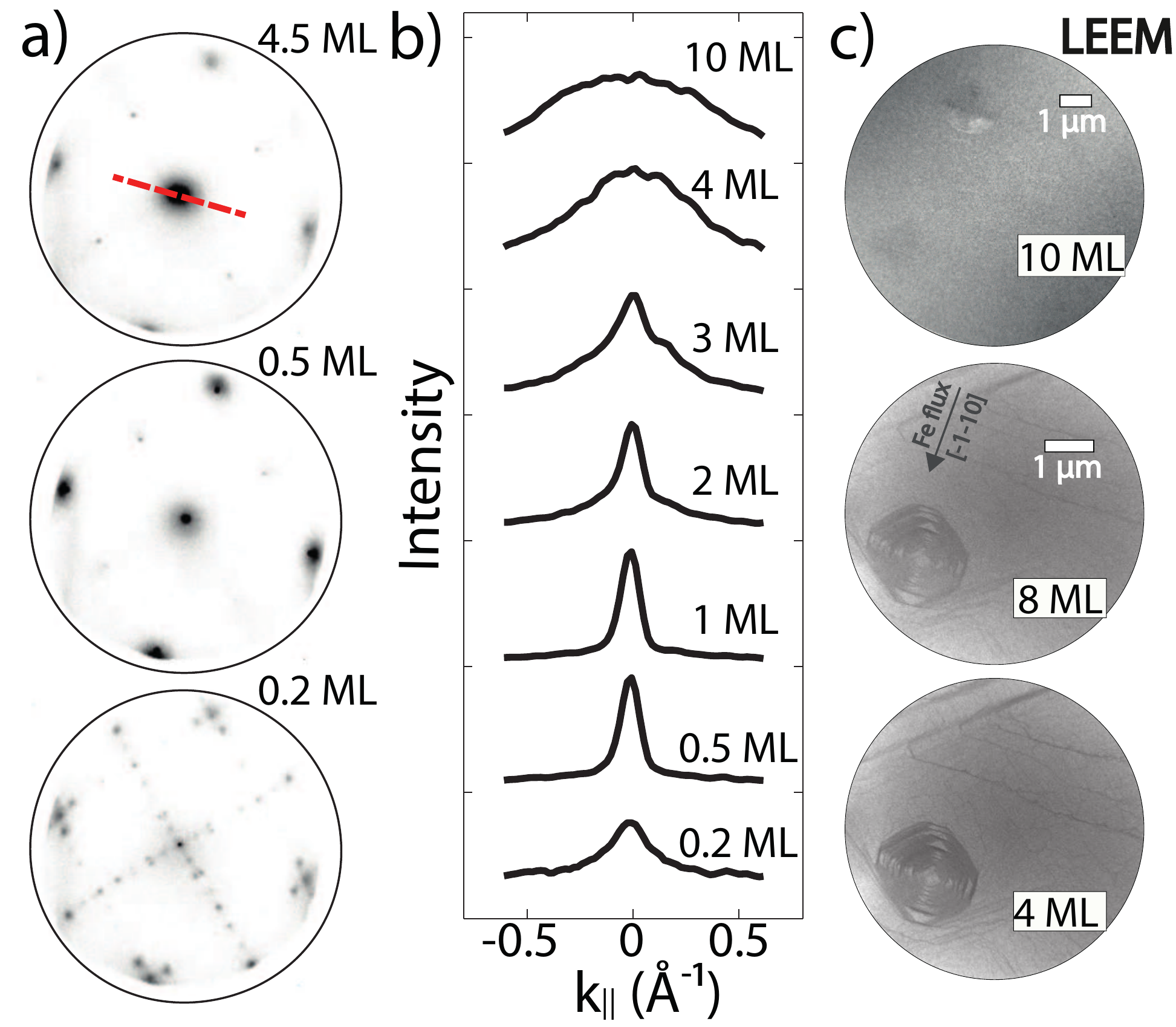}

\caption{\label{fig:fig2} a) The LEED patterns at different deposition times taken during growth at 50~eV kinetic energy.  b) The angular profiles of the LEED~(0,0) spot for different Fe coverage. c) LEEM images taken at RT using 12~eV electron kinetic energy at different Fe film thicknesses. For 10~ML coverage the steps propagating from the Au(001) substrate are no longer visible.}
\end{figure}

\begin{figure}
\includegraphics[width=\columnwidth]{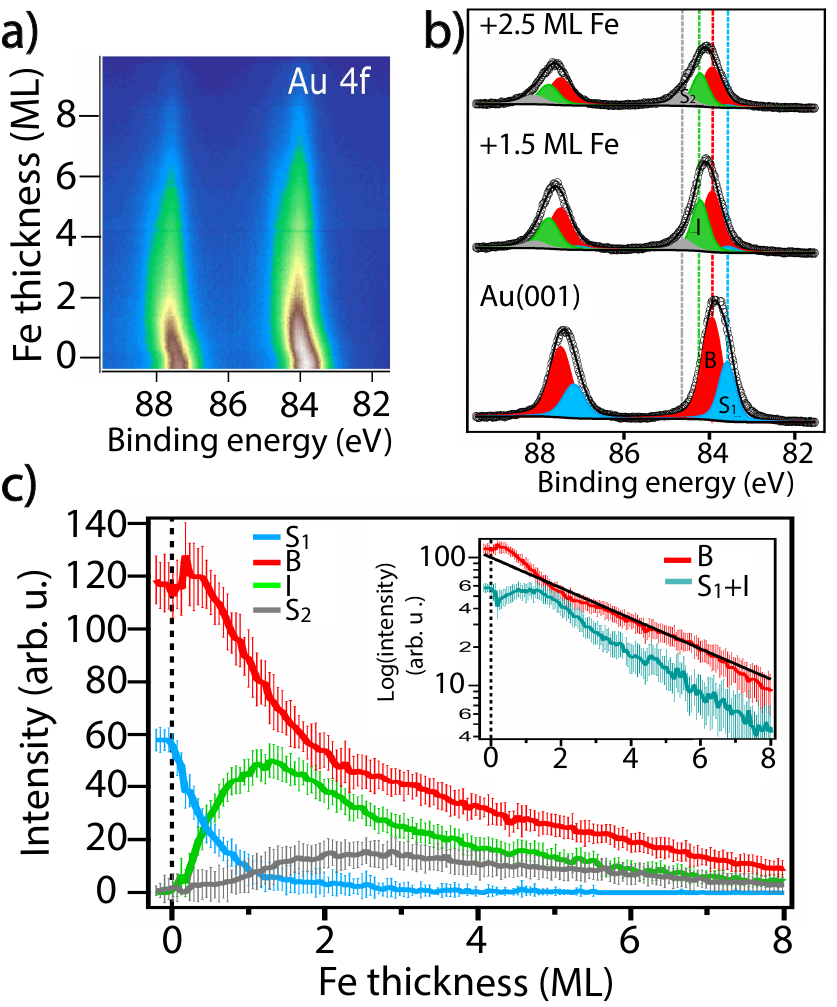}
\caption{\label{fig:features} a) A merge of XPS spectra of the Au 4f doublet taken during the deposition of Fe at RT for different film thicknesses. b) Fit of the XPS spectra from pure Au(001) surface and two chosen thicknesses of the Fe film. The components of the fit: B (Au bulk), I (Fe/Au interface), S$_{1}$ (Au surface), and S$_{2}$ (Au/Fe) are described in the text. c) Intensity of the components of the fit with the corresponding standard deviation error bars plotted versus Fe film thickness. Inset shows the component B and the sum of the components S$_{1}$ and I plotted on the logarithmic scale. Black line represents theoretical attenuation of Au signal by Fe layer for IMFP equal to 5.22~\AA.}
\end{figure}

To characterize the evolution of the surface composition during growth, we performed a complementary XPS study. The Au 4f core level was monitored by imaging the analyzer dispersive plane on the detector during Fe deposition. The photoemission spectra were obtained by collecting intensity profiles along the dispersion direction. The spectra of the Au 4f core level acquired using photon energy of 250~eV as a function of the Fe coverage are presented in Fig. 3 (a). Clearly, the intensity of the Au 4f peak diminishes with the Fe thickness. After the deposition of 10~ML of Fe the total intensity of the Au 4f peak reduces to 5 \% of the initial value, which is in a very good agreement with the theoretically predicted attenuation, taking into account the inelastic mean free path (IMFP) of $\lambda = 5.22$~\AA \space for electrons with kinetic energy 159~eV. A higher intensity of the Au 4f peak would be expected for Au present on the surface of the Fe film. What can also be seen, is the apparent shift of the center of the Au 4f doublet towards higher binding energies at the very beginning of the evaporation process.
\par
To quantify these observations, after the subtraction of a Shirley background, all measured spectra were consistently fitted with a set of components of the same Doniach-Sunjic peak shape, characterized by an asymmetry parameter $ \alpha  = 0.01$.  The exemplary fits are presented in Fig. 3 (b). A satisfactory fit for the entire series was obtained by introducing four components with fixed binding energies: S$_{1}$: 83.6~eV, B: 83.9~eV, I: 84.2~eV, and S$_{2}$: 84.6~eV. Their positions are marked in Fig. 3 (b) with dashed vertical lines. The FWHM of the components ranged between 0.45 and 0.6~eV. 
\par

The area under each of the components versus Fe thickness is plotted in Fig. 3 (c). Error bars represent standard deviation calculated using the Monte Carlo method, as implemented in the CasaXPS software.\cite{Fairley2005} The component B can be unambiguously identified as stemming from the bulk of the Au crystal, as it is the most intense signal when measured on the clean Au(001) substrate. The additional component present before the evaporation (component S$_{1}$) originates from the reconstructed surface of the Au(001) crystal (Fig. 1). The binding energy $E_B$ of the surface component is shifted with respect to the bulk $E_B$ by -0.3~eV, which is in a very good agreement with the values reported earlier.\cite{Heimann1981} The intensity of the surface component S$_{1}$ ($E_B=83.6$~eV) decreases immediately at the beginning of the evaporation down to almost zero (taking into account the uncertainty values) when 1~ML of Fe is reached (Fig. 3 (c), blue curve). This indicates that the second Fe ML starts to grow after the first one is completed. Simultaneously, component I ($E_B=84.2$~eV) grows, reaching maximum intensity slightly above 1~ML coverage and decaying exponentially for higher coverage. We interpret component I as originating from the interface between the Fe film and the surface of the Au crystal. Starting at approximately 1~ML, another small component at a higher binding energy appears (component S$_{2}$,  $E_B=84.6$~eV). For components I and S$_{2}$, the binding energy shift with respect to the bulk component equals to +0.3~eV and +0.7~eV, respectively. Such binding energy shifts can be attributed to the Fe-Au bonding (Ref. \onlinecite{Naitabdi2009} and references therein). We interpret component S$_{2}$ as originating from the Au atoms present on top of the Fe surface. The intensity of the component S$_{2}$ stays constant up to the coverage of approximately 3~ML. At this point, it contributes 18 \% of the total spectral intensity, indicating sub-monolayer coverage of the corresponding Au overlayer atoms. Above 3~ML, the component S$_{2}$ also starts to gradually decay. Therefore, we conclude that beyond 3~ML coverage the Au overlayer on the surface starts to be covered by Fe atoms. However, a non-zero S$_{2}$ component remains present at 10~ML coverage indicating intermixing of the Au atoms of the overlayer into the layers of the deposited Fe film. In the inset of Fig. 3 (c), the intensity of the bulk (B) component and the sum of the surface and interface components (S$_{1}$ + I) are shown on a logarithmic scale. Additionally, a black solid line, which represents the theoretical attenuation of the Au signal by the Fe film taking into account an IMFP of 5.22~\AA, is also plotted. We see that the linear decrease of both curves (B and S$_{1}$+I) is well approximated by the theoretical prediction, which further justifies the applied model. 
\par

%%%%%%%%%%%%%%%%%%%%%%%%%%%%%%%%%%%%%%%%%%%%%%%%%%%%%%%%%%%%%%%%%%%%%%%%%%%%%%%%%%%%%%%%%%%%%%
%%%%%%%%%%%%%%%%%%%%%%%%%%%%%%%%%%%%%%%%%%%%%%%%%%%%%%%%%%%%%%%%%%%%%%%%%%%%%%%%%%%%%%%%%%%%%%
\section{A\MakeLowercase{u} SEGREGATION DURING ANNEALING}
\label{Sec3}
The 10~ML Fe film was annealed to 300~$^{\circ}$C. After the annealing procedure and cooling down to RT, we obtained XPEEM images at different binding energies using 150~eV photon energy (Fig. 4 (a)). We observed bright regions in the XPEEM images taken at the Au 4f$_{7/2}$ binding energy peak, which appeared dark in the image taken at Fe 3p peak.  The XPS spectra confirm the difference in the intensity of the Au 4f peaks and the Fe 3p depending on the position on the surface of the annealed Fe film. We attribute the XPS intensity difference to a non-homogeneous thickness of the segregated Au on the surface of the Fe film.
\par
The quantification of the thickness of the Au overlayer was performed taking into account the intensity ratio of bulk Au 4f and Fe 3p lines, which for the photon energy of 150~eV equals to $I_{0}(Au)/I_{0}(Fe) = 0.95$. This value was calculated according to Wagner \textit{et al.}.\cite{Wagner1981} Selected-area XPS spectra resulted in the intensity ratios of $I(Au)/I(Fe) = 1.03$ and $I(Au)/I(Fe) = 1.79$ for the different areas marked in the Fig. 4 a). This translates into 1.6~ML and 2.3~ML of Au overlayer, respectively. In this calculation we used an Au-Fe interlayer spacing of 1.75~\AA \space and took into account the substrate contribution to the Au peak (5\% of the bulk intensity). 
\par
\begin{figure}
	\includegraphics[width=\columnwidth]{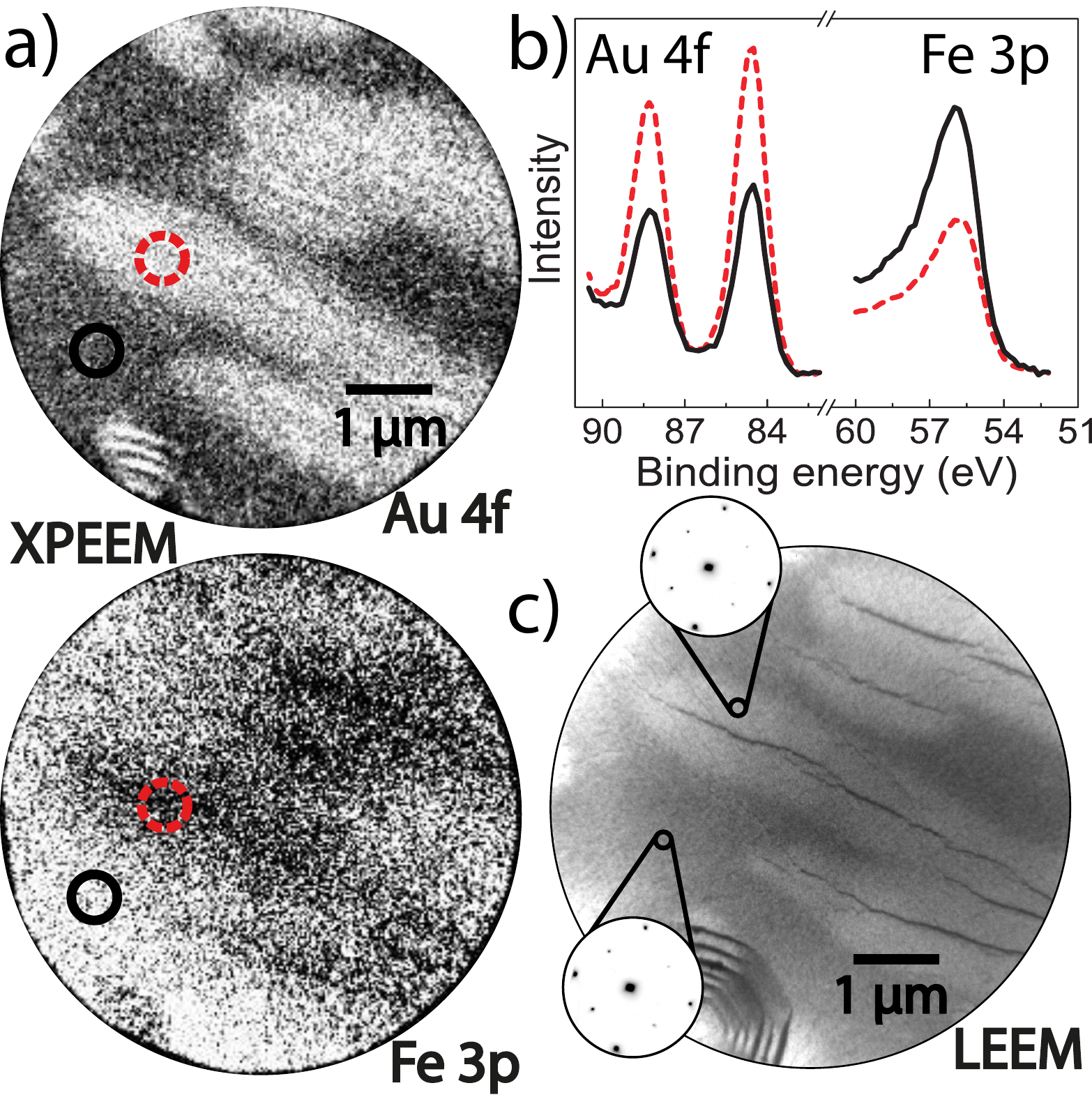}
	\caption{\label{fig:features} a) XPEEM images taken at the Au 4f$_{7/2}$ and Fe 3p binding energy. Intensity was normalized to the background. b) XPS spectra obtained from different regions on the surface marked in a). c) A LEEM image of the same area as in (a) where the cracks can be observed with the \textmu -LEED images taken from the Au-rich and the Fe-rich areas shown in the insets.}
\end{figure}
\begin{figure*} 
	\centering
	\includegraphics[width=0.8\textwidth]{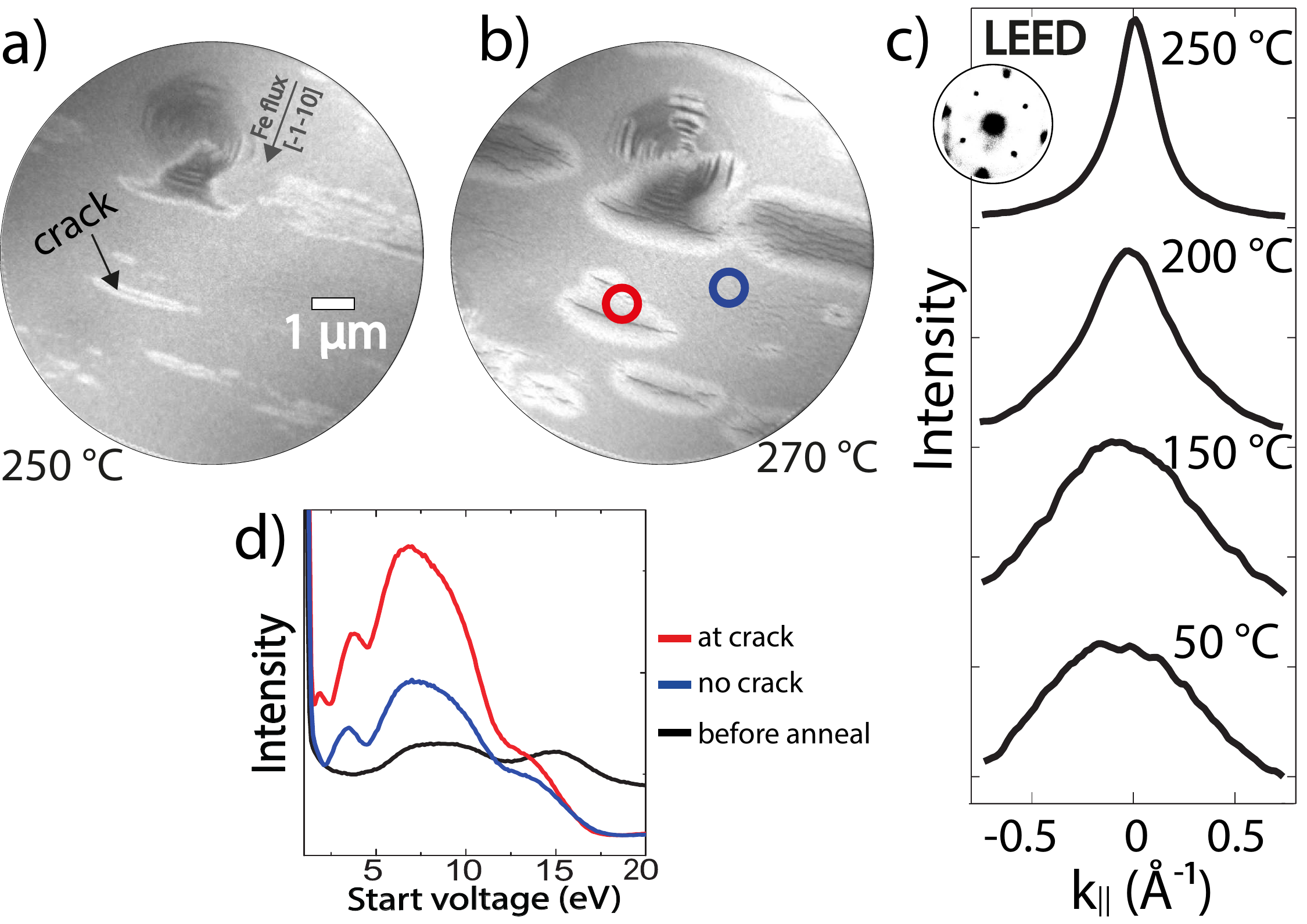}
	\caption{\label{fig:all_sbs}  LEEM images obtained at electron kinetic energy of 12~eV during annealing at 250~$^{\circ}$C a) and 270~$^{\circ}$C b). The direction of Fe deposition is marked with an arrow in a). The cracks were observed to have a dark contrast and the bright area surrounding the cracks increases at higher temperatures. c) The 0-order spot profiles of the LEED pattern (inset) recorded at the electron energy of 50~eV during annealing at temperatures marked in the image. e) I-V LEEM curves measured for kinetic energies between 0 and 20~eV before (black curve) and after the annealing procedure (blue and red curve). The selected area on the surface is marked in b) (blue and red circle).}
\end{figure*}
In the LEEM image of the same region acquired using electron kinetic energy of 12~eV after annealing (Fig. 4 (c)) we observed 1-5~\textmu m long \textit{cracks} with bright surrounding areas, which corresponded to the regions of different contrast observed in XPEEM (Fig. 4 (a)). In the subsequent experiment, the annealing step was monitored with LEEM (Fig. 5). In LEEM images obtained during annealing we identified the opening of 1-5~\textmu m long \textit{cracks} in the Fe film at temperatures above 190~$^{\circ}$C (Fig. 5 (a and b)). At higher temperatures bright areas around the cracks appear and increase in size. Intensity-voltage LEEM (I-V LEEM) was acquired before and after the annealing process by varying the electron kinetic energy from 0 to 20~eV (Fig. 5 (d)). The intensity profiles from selected areas (of $\sim $0.5 micrometer diameter) in the image taken before annealing (black line) and after annealing (close to the crack (red line) and away from the crack (blue line)) are plotted versus the kinetic energy. The shape of I-V LEEM curves can be qualitatively explained by comparing them to the unoccupied states of the electronic band structure of the material.\cite{Strocov1999, Bauer1998} We observe clear changes in the intensity modulation of the I-V curve taken before and after annealing, indicating changes in the surface band structure. By selecting areas of the image we were able to analyze local changes in the I-V LEEM curves. The curves taken after annealing, close and away from the crack, are also distinct. When selecting only the area close to the cracks, we observe two peaks between 0 eV and 6.5~eV, while for the area away from the crack only one peak is visible. We attribute the changes in the I-V curves to the difference in the thickness of the segregated Au layer. However, for a full understanding of the I-V LEEM curves dedicated theoretical calculations would be mandatory.\par

During annealing the LEED pattern was also monitored. The width of the 0-order spot profile decreases with increasing temperature above $\sim $ 120~$^{\circ}$C, as can be seen in Fig. 5 (c). Annealing to 250~$^{\circ}$C and above leads to a Lorentzian-shaped angular profile, with no background intensity. At the same time the sharpness of the first order diffraction spots significantly improves. This could suggest a substantial reduction of the surface roughness, but it may also be related to an increased amount of Au atoms on the surface of the Fe film. With areal selection on the surface we observed the typical bcc Fe LEED pattern in both, the Fe and the Au rich regions (insets in Fig. 4 (c)). No changes in the LEED pattern were observed during cooling to RT.

\par
Jiang \textit{et al.} \cite{Jiang1993} showed that the temperature at which the Au atoms intermix by place exchange in the Fe film surface depends on the Fe coverage. For a 5~ML Fe film on Au the authors observed an increase of intermixing of Au in the surface at annealing temperatures above 200~$^{\circ}$C. This agrees with the temperature of the formation of the cracks in our study, indicating that the thickness of Fe in the central region of the crack is reduced. Enhanced segregation of Au to the surface followed by the surface diffusion in the directions normal to the crack is most probably induced by the low surface free energy of Au, smaller almost by a factor of 3 compared to Fe.\cite{Bonell2013} Similar observations were made by Schmid \textit{et al.} \cite{Schmid1993} for the Co thin films grown on Cu(100).
\par

From the LEEM images (taken approximately every 8~s) during annealing the growth rate of the brighter Au-rich regions around cracks was determined to be in the order of magnitude of 10\textsuperscript{-11} cm\textsuperscript{2}/s at temperatures between 250 and 300~$^{\circ}$C. This is consistent with the surface diffusion coefficient D (according to the Arrhenius law $ D=D_{0}\cdot exp(-E_{a}/k_{B}T) $ with $D_{0}$ being the pre-exponential factor and $E_{a}$ the diffusion activation energy \cite{Liu1991}) for self-diffusion of Au by the hopping mechanism calculated by Liu \textit{et al.}\cite{Liu1991} using embedded atom model and by Sanders and DePristo.\cite{Sanders1992} This supports our conclusion that the growth of the Au-rich regions is due to the surface diffusion of Au atoms after segregation to the surface through the channels of reduced Fe thickness. However, the segregation through the cracks is possibly not the only mechanism of Au segregation. A different mechanism which does not require the formation of cracks or deformations (e.g. in the study of Zdyb \textit{et al.}\cite{Zdyb2009} they do not observe formation of cracks in the LEEM images) may be the leading segregation mechanism in the regions, where we obtained a 1.6~ML Au overlayer.
\par

Interestingly, we observed the cracks in the Fe film almost solely along the direction close to Fe[1$\bar{1}$0]. This, in principle, may be related to a significant miscut of the Au(001) single crystal or a strain from the sample holder. However, because we did not observe indications for such interpretation in any other measurements, we propose another possible explanation of the unidirectional orientation of the observed cracks and the localized increase of Au atoms on the surface. The near-grazing angle geometry can cause shadowing effects during the deposition. It has been shown that the epitaxial growth of metal thin films in the oblique incidence geometry results in increased surface roughness and formation of mounds, the shape of which depends on the deposition angle, temperature and the thickness of the film.\cite{Dijken2000, Shim2008, Rabbering2010} For deposition angles beyond 50$^{\circ}$ off the surface normal a phenomenon named \textit{steering} was suggested by Dijken \textit{et al.} \cite{Dijken2000}.  It was used to explain the increased deposition flux on top of surface protrusions observed for Cu/Cu(001).\cite{Dijken2000, Rabbering2010} Due to the islands and steps present on the substrate surface, the attractive potential between the incident atoms and the substrate is distorted and causes changes of the trajectories of the incident atoms. As a result, the incident flux is increased on top of protruding terraces and decreased behind descending steps. Interestingly, the simulations performed by Dijken \textit{et al.}\cite{Dijken2000} predicted a larger area of reduced flux due to steering behind steps compared to the classical shadowing effect. Thus, the high step and step bunch concentration on the clean Au single crystal surface (visible in Fig. 1(b)) can lead to Fe poor regions behind descending steps on the Au surface in the direction Fe[1$\bar{1}$0], i.e. perpendicular to the deposition direction. %, see Fig. 6 (a and b). 
In addition to classical shadowing, the steering effect may lead to an asymmetry in the deposition flux on top of the terrace and behind a step. Thus, the short distance between many steps which form a step bunch %(sketched in Fig. 6 (a)) 
can result in an increased size of the low Fe coverage area. At an increased temperature during annealing the diffusion of Au atoms %is 
would be enhanced at the Fe-poor regions, which would act as a channel for segregation of Au to the surface. Verification of such a scenario requires further experiments, which may be inspired by this work.
\par
\section{Conclusions}

A new phenomenon has been observed during the preparation of a thin Fe film on a Au(001) single crystal which leads to Au atoms crawling on top of Fe surface though the cracks which form in the 10~ML Fe films during annealing above 200~$^{\circ}$C. This shines a new light on the debate on whether a monolayer of Au is always formed on top of Fe film, where several previous studies have made contradicting conclusions.\cite{Bader1998,Begley1993,Blum1999,He1993,Kellar1998,Rossi1995} In images acquired using LEEM and PEEM we found that 1-5~\textmu m long cracks open in the Fe film when heated above 190~$^{\circ}$C. The microscopic images and XPS spectra showed an increased segregation of Au through the cracks at temperatures above 200~$^{\circ}$C. The thickness of the overlayer was found to be 1.6~ML away from the cracks and 2.3~ML in the regions on the film close to the cracks. Compared to below-ML thickness of Au on top of Fe film identified from XPS spectra obtained before increasing the temperature we conclude that the annealing step was crucial for the increased segregation of Au. 
\par
The non-uniform Au segregation as well as the possible intermixing at the interface of this system locally influence the strength of the interfacial Rashba spin-orbit interaction, which is interpreted as one of the origins of the spin-orbit torque. Moreover, the observed unidirectional cracks acting as a source of increased Au segregation during annealing may add to the asymmetry of the conductance measured in a Fe(10~ML)/Au(001) based device. Therefore, a thin Fe film grown on Au at RT is not the favorable model system to study the current-driven magnetization at a crystalline ferromagnet/heavy-metal interface. For studies of the spin-orbit torque phenomena we suggest to reverse the system and choose a suitable single crystalline insulating substrate. On the other hand, the observed segregation of Au in an Fe(10~ML)/Au(001) system might be reduced by annealing at temperatures below 190~$^{\circ}$C. The results of our experiment prove the necessity to employ spatially-resolved methods in studies of growth, also for the layered epitaxial metal-metal heterostructures.
\par 
%%%%%%%%%%%%%%%%%%%%%%%%%%%%%%%%%%%%%%%%%%%%%%%%%%%%%%%%%%%%%%%%%%%%%%%%%%%%%%%%%%%%%%%%%%%%%%
%%%%%%%%%%%%%%%%%%%%%%%%%%%%%%%%%%%%%%%%%%%%%%%%%%%%%%%%%%%%%%%%%%%%%%%%%%%%%%%%%%%%%%%%%%%%%%
\section{Acknowledgments}
The authors would like to thank A. Locatelli and O. Mentes for their support at the Nanospectroscopy beamline and C. Schmitz and M. Giesen for fruitful discussions. This work was supported by The Initiative and Networking Fund of the Helmholtz Association. 
\par
%%%%%%%%%%%%%%%%%%%%%%%%%%%%%%%%%%%%%%%%%%%%%%%%%%%%%%%%%%%%%%%%%%%%%%%%%%%%%%%%%%%%%%%%%%%%%%
%%%%%%%%%%%%%%%%%%%%%%%%%%%%%%%%%%%%%%%%%%%%%%%%%%%%%%%%%%%%%%%%%%%%%%%%%%%%%%%%%%%%%%%%%%%%%%

\bibliography{LP-Mendeley3}% Produces the bibliography via BibTeX.
\end{document}